\documentclass[
aps,prm,%
preprint,
superscriptaddress
]{revtex4-2}
\usepackage{color}
\usepackage{graphicx}
\usepackage{lineno}
\usepackage{gensymb}
\usepackage{textcomp}
\usepackage{booktabs}
\usepackage{multirow}
\usepackage{amsmath}
\usepackage[dvipdfmx,colorlinks=true]{hyperref}
\hypersetup{
 linkcolor=blue,
 citecolor=blue,
 urlcolor=blue
 }

\begin{document}

\title{Large Spin Hall Effect in High-Entropy Alloy/CoFeB Bilayers}
\author{Takahide Kubota}
\email[]{takahide.kubota@tohoku.ac.jp}
\affiliation{Department of Advanced Spintronics Medical Engineering, Tohoku University, Sendai 980-0845, Japan}
\affiliation{Institute for Materials Research, Tohoku University, Sendai 980-8577, Japan}
\author{Kazuya Z. Suzuki}
\affiliation{Advanced Science Research Center, Japan Atomic Energy Agency, Tokai 319-1195 Japan}
\affiliation{Advanced Institute for Materials Research, Tohoku University, Sendai 980-8577, Japan}
\author{Yoshiyuki Hirayama}
\affiliation{Samsung Device Solutions R\&D Japan, Yokohama 230-0027, Japan}
\author{Shigeki Takahashi}
\affiliation{Samsung Device Solutions R\&D Japan, Yokohama 230-0027, Japan}
\author{Koki Takanashi}
\email[]{takanashi.koki@jaea.go.jp}
\affiliation{Advanced Science Research Center, Japan Atomic Energy Agency, Tokai 319-1195 Japan}
\affiliation{Advanced Institute for Materials Research, Tohoku University, Sendai 980-8577, Japan}
\affiliation{Institute for Materials Research, Tohoku University, Sendai 980-8577, Japan}
%
%
\begin{abstract}
High-entropy alloys (HEAs) exhibit various physical properties, such as high microhardness for structured materials and high efficiency for catalysis. These features are recognized as a cocktail effect of five or more elements that stabilize a single-phase solid solution due to a high configurational entropy.
HEAs may also exhibit short-range orders and microdistorshons, which cause the local symmetry breaking of systems. Systems with local symmetry breaking are of interest for spin-dependent transport, such as for spin Hall effects.
In this study, sputtered-film samples of an HEA, Mn--Nb--Mo--Ta--W, and related alloys, were fabricated; these films were layered with ferromagnetic CoFeB.
All samples exhibited x-ray diffraction peaks originating only from a body-centered cubic (bcc) phase, and transmission electron microscopy images indicated the absence of secondary phases and uniform elemental distributions.
The spin Hall magnetoresistance (SMR) was investigated, and clear resistance changes were observed in all the samples. Quantitative analysis of SMR revealed that the spin Hall angle ($\theta_\mathrm{SH}$) was 0.12 $\pm$ 0.002 and 0.14 $\pm$ 0.037 for the HEA and Nb--Mo--Ta--W alloy (medium-entropy alloy (MEA)), respectively.
The $\theta_\mathrm{SH}$s of the HEA and MEA were comparable to that of Pt which is a typical heavy element with relatively large $\theta_\mathrm{SH}$.
The newly developed HEA and MEA films are attractive spin Hall materials with the bcc phase and are suitable for spintronic applications using magnetic tunnel junctions with CoFeB and an MgO barrier.
\end{abstract}
\date{\today}%
\maketitle

\section{Introduction}
\label{sec:intro}
High-entropy alloys (HEAs) are qualitatively defined as alloys that consist of at least five metallic elements of roughly equimolar concentrations and exhibit a crystallin single-phase solid solution. They have garnered considerable interest due to their various practical applications, such as structured materials exhibiting high microhardness, and catalytic efficiency~\cite{Yeh2004,Cantor2004,Tsai2014,Miracle2017,Sun2021,Fujita2023}.
HEAs can also be quantitatively defined using the configurational entropy of mixing for an ideal solid solution, $S^\mathrm{mix, ideal} = -R\sum_{i} x_i ln(x_i)$, where $R$ is the gas constant and $x_i$ is the atomic fraction of element $i$. Under this, alloys with multiple principal elements showing $S^\mathrm{mix, ideal} \ge 1.61R$ are defined as HEAs.  Medium entropy alloys (MEAs) are alloys with $S^\mathrm{mix, ideal}$ raging between 0.69$R$ and 1.61$R$
. Although the quantitative definition using $S^\mathrm{mix, ideal}$ clearly classifies materials, the qualitative definition is also commonly used due to its simplicity\cite{Yeh2004,Cantor2004,Tsai2014,Miracle2017,Sun2021,Fujita2023}. 
\\
MEAs and HEAs exhibit short-range order and microdistortion originating from the single phase formed by multiple elements with various atomic radii\cite{Yeh2004,Miracle2017,Furuya2024}.
Such a broken symmetry has garnered interest in terms of spin transport, \textit{e.g.}, in spin Hall effects, which are spin-dependent transport phenomena of conduction electrons in systems exhibiting symmetry breaking\cite{Dyakonov1971,Hirsch1999,Valenzuela2006,Hoffmann2013,Morota2011,Sinova2015,Hirohata2020}.
The spin Hall effect involves the conversion of a charge current into a spin current that causes spin accumulation and can manipulate local magnetic moments via the spin-orbit torque\cite{Ando2008,Miron2011,Liu2012,Saidaoui2016,Manchon2019,Song2021,Shao2021}. Thus, the enhancement of factors indicating charge-to-spin conversion efficiency, such as spin Hall angle ($\theta_\mathrm{H}$), is a crucial topic in spintronics, which can be explored by developing new materials showing the spin Hall effects.
Because materials with large atomic numbers ($Z$) are associated with pronounced spin Hall effect, which intrinsically correlates with spin--orbit interaction proportional to $Z$, many studies have been conducted on heavy transition-metal elements such as Pt, Au, Pd, Hf, Ta, W, and Re\cite{Kimura2007,Seki2008,Ando2010,Morota2011,Liu2012,Pai2012,Liu2015,Magni2022} binary alloys and multilayers thereof; and light 3d elements with low doping of these heavy elements\cite{Niimi2012,Kim2020,Masuda2020,Saito2020,Coester2021}.
Although these previous studies revealed pronounced spin Hall effects using heavy metals, HEAs and MEAs have been insufficiently studied.
One pioneering paper\cite{Chen2017} reported the demonstration of magnetization reversal by spin--orbit torque in a layered structure using an HEA, Ta--Nb--Hf--Zr--Ti; however, pure Hf and Ta were located at the interface between the HEA and a ferromagnetic CoFeB layer. In addition, the Ta--Nb--Hf--Zr--Ti layer exhibited a nano-crystalline amorphous-like phase; however, its spin Hall efficiency was not quantitatively evaluated\cite{Chen2017}.
In this study, the spin Hall effects of the Mn--Nb--Mo--Ta--W HEA and Nb--Mo--Ta--W MEA were investigated. The former exhibits a stable bcc phase and high microhardness\cite{Senkov2011}. Adding V to it results in V--Nb--Mo--Ta--W, is a well-known HEA; however, in this study, Mn, a magnetic element, was used instead of V to obtain an enhanced spin scattering effect.
To investigate the spin Hall effects in HEA and MEA films, spin Hall magnetoresistance (SMR)\cite{Nakayama2013,Kim2016} of HEA (or MEA)/CoFeB layered structures was measured.

\section{Experimental Procedures}
\label{sec:exp}
\begin{table*}
\begin{center}
\caption{
Film compositions and the configurational entropy of an ideal solid solution, $S^\mathrm{mix, ideal}$ of the high-entropy alloy (HEA) and medium-entropy alloys (MEAs) in this study. These notations are used in the main text and figures.
}
\label{t:samples}
\begin{ruledtabular}
\begin{tabular}{c ccccc c c}
\multirow{2}{*}{Notation} &\multirow{2}{*}{Stoichiometry} & \multicolumn{5}{c}{Compositions (at.\%)} & \multirow{2}{*}{$S^\mathrm{mix,ideal}$}\\
 \cline{3-7}
 & & Nb & Mo & Ta & W & Mn &\\
\hline
MEA1 & Nb$_{1.1}$Mo$_{0.9}$ & 53.7 & 46.3 & -- & -- & -- & 0.69$R$\\
MEA2 & Nb$_{0.3}$Mo$_{0.2}$Ta$_{1.6}$W$_{1.9}$ & 6.1 & 6.0 & 40.8 & 47.0 & -- & 1.06$R$\\
MEA3 & Nb$_{1.4}$Mo$_{1.3}$Ta$_{0.6}$W$_{0.7}$ & 35.8 & 31.8 & 15.2 & 17.2 & -- & 1.32$R$\\
MEA4 & Nb$_{1.2}$Mo$_{1.1}$Ta$_{0.8}$W$_{0.9}$ & 30.7 & 26.8 & 20.1 & 22.4 & -- & 1.37$R$\\
HEA & Mn$_{0.9}$Nb$_{1.2}$Mo$_{1.1}$Ta$_{0.8}$W$_{0.9}$ & 18.7 & 17.3 & 24.5 & 21.4 & 18.1 & 1.60$R$\\
\end{tabular}
\end{ruledtabular}
\end{center}
\end{table*}
Layered film samples were fabricated onto thermally oxidized silicon substrates using an ultrahigh-vacuum magnetron sputtering system (base pressure of the sputtering chamber was 2 $\times 10^{(-7)}$ Pa or lower).
Film compositions were controlled by adjusting the power supplies for cosputtering process using Nb--Mo alloy, Ta, W, and Mn targets. The compositions of the fabricated films were then evaluated using an electron probe micro analyzer. The film compositions and $S^\mathrm{mix, ideal}$ of HEA (or MEA) layers are summarized in Table \ref{t:samples}.
Although $S^\mathrm{mix, ideal}$ is slightly smaller than the lower boundary of the quantitative definition of HEAs, Mn$_{0.9}$Nb$_{1.2}$Mo$_{1.1}$Ta$_{0.8}$W$_{0.9}$ was considered as an HEA based on the qualitative definition. Other alloys are denoted as MEA1--MEA4 (Table \ref{t:samples}).
The stacking structure was substrate$/$HEA (or MEA) layer $t$: 2--7 nm$/$CoFeB 1.5 nm$/$MgO 1.0 nm$/$Ta 1.0 nm. All layers were deposited at room temperature. Some samples were annealed at 350 \degree C in a vacuum furnace for 1 hour, and the as-deposited samples were compared with the annealed samples to clarify the stability of the crystalline phases of the samples.
In the layered structure, the HEA (or MEA) layer is a spin current source and the CoFeB layer behaves as a spin absorber. The MgO$/$Ta layers are protection layers, in which Ta was naturally oxidized  and did not contribute to spin transport.
The crystal structures were characterized using X-ray diffraction with Cu-K$_\alpha$ radiation and scanning transmission electron microscopy (TEM).
SMR was measured by means of the four-probe technique at 300 K under an applied magnetic field of 7 T.

\section{Results and Discussion}
\label{sec:results}
\begin{figure}
\begin{center}
\includegraphics[clip,scale=0.8]{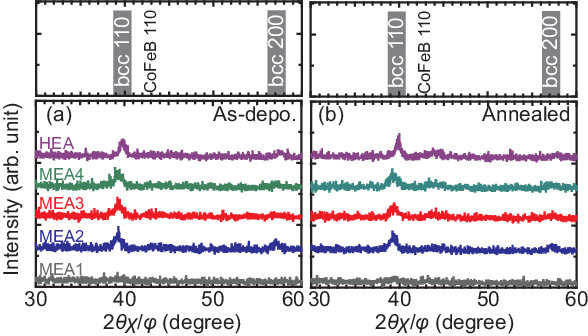}
\caption{In-plane XRD patterns of Mn--Nb--Mo--Ta--W 5 nm/CoFeB 1.5 nm bilayer samples. The top panels indicate approximate positions of 1~1~0 and 2~0~0 diffraction peaks for a body-centered cubic (bcc) phase and a 1~1~0 peak for the CoFeB layer.}
\label{fig:xrd}
\end{center}
\end{figure}
The in-plane XRD patterns are shown in Figs. \ref{fig:xrd}(a) and \ref{fig:xrd}(b) for the as-deposited and annealed samples, respectively.
All samples clearly exhibited 1~1~0 and 2~0~0 diffraction peaks of a bcc phase of HEA and MEA layers, except for Nb$_{1.1}$Mo$_{0.9}$ sample that showed very weak 1~1~0 diffraction around 39\degree. Broad diffraction peaks were observed around 43\degree for the annealed samples, which possibly originated from the bcc phase of the CoFeB layer.
The diffraction peaks of HEAs and MEAs did not change considerably before and after annealing. The absence of additional peaks suggests that the samples are free from crystalline secondary phases.
\begin{figure}
\begin{center}
\includegraphics[clip,scale=0.7]{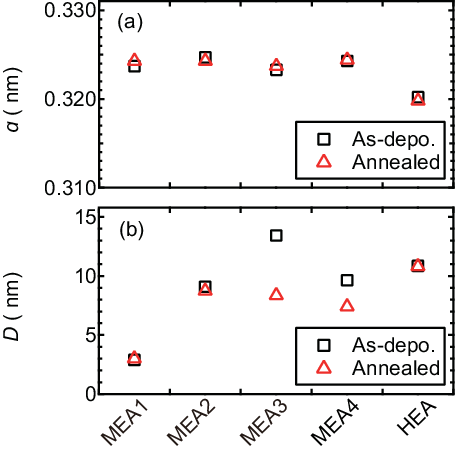}
\caption{Composition dependence of (a) lattice constants and (b) grain sizes of 5-nm-thick Mn--Nb--Mo--Ta--W layers in the bilayer samples. The black squares and red triangles represent data points for as-deposited and annealed samples, respectively.}
\label{fig:lattice}
\end{center}
\end{figure}
Lattice constants ($a$) and grain sizes of the HEA and MEA layers were evaluated for the as-deposited and annealed samples based on the diffraction angles and width of the 1~1~0 peaks (Figs. \ref{fig:lattice}(a) and \ref{fig:lattice}(b)). Scherrer's equation was used to determine the grain sizes, as follows~\cite{Scherrer1918},
\begin{equation}
\label{eq:Scherrer}
D = \frac{K\lambda_{\mathrm{Cu}K_\alpha}}{B\cos\theta},
\end{equation}
where $D$ is the grain size, $K$ is the Scherrer constant (0.94), $\lambda_{\mathrm{Cu}K_\alpha}$ is the X-ray wavelength for Cu-K$_\alpha$ radiation, $\theta$ is the diffraction angle of a peak in 2$\theta$ scan, and $B$ is the full-width at half-maximum of the peak.
The lattice constants were about 0.325 and 0.320 nm for the MEAs and HEA respectively. This indicated that the as-deposited and annealed samples were almost similar; these findings correlated with the XRD patterns.
The evaluated lattice constants of the MEAs were consistent with the average values of the lattice constants derived based on the atomic fractions of constituent elements in the MEAs.\cite{Vegard1921,Denton1991,Wang2019}, \textit{i.e.}, they are in between those of Nb (0.330 nm), Mo (0.315 nm), Ta (0.331 nm), and W (0.316 nm)\cite{Hermann2011}. Although no information is available for pure Mn showing a bcc phase, a relatively small van der Waals atomic radius of Mn (0.197 nm)\cite{PubChem2025} possibly reduced the lattice constant of the HEA. Note that the atomic radii of Nb, Mo, Ta, and W are 0.207, 0.209, 0.217, and 0.210 nm, respectively\cite{PubChem2025} .
MEA1 exhibited a relatively small value of $D$ of $\sim$3 nm, whereas other samples exhibited relative large $D$ values of 7--14 nm. 
Considering that sputtering deposition is a kind of quenching crystallization process from the high-energy sputtered atoms, the composition dependence of $D$ for the as-deposited samples is difficult to discuss here. Meanwhile, the annealed samples were considered to be in a stable state. Based on these considerations, $D$ values are relatively large for samples with $S^\mathrm{mix,ideal}$ greater than 1; this suggests that the crystallized grains were stabilized by the entropy effect.
\begin{figure}
\begin{center}
\includegraphics[clip,scale=0.8]{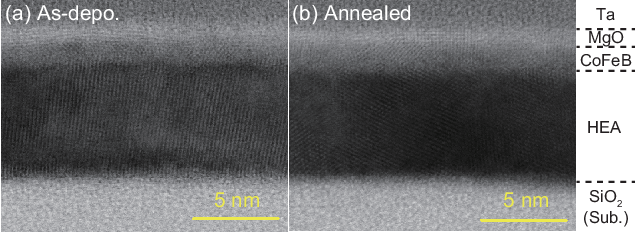}
\caption{Bright-field cross-sectional transmission electron microscopy images of (a) as-deposited and (b) annealed samples. The nominal layer thickness is 7 nm for the Mn--Nb--Mo--Ta--W layer.}
\label{fig:TEM}
\end{center}
\end{figure}
To investigate the microstructures of the HEA samples, the cross-sectional TEM was performed. The corresponding bright-field images are shown in Figs. \ref{fig:TEM}(a) and \ref{fig:TEM}(b) for the as-deposited and annealed samples, respectively. For both images showed layered structures according to the designed stacking sequence, and a crystalline single phase was confirmed for the HEA layers.
\begin{figure*}
\begin{center}
\includegraphics[clip,scale=0.8]{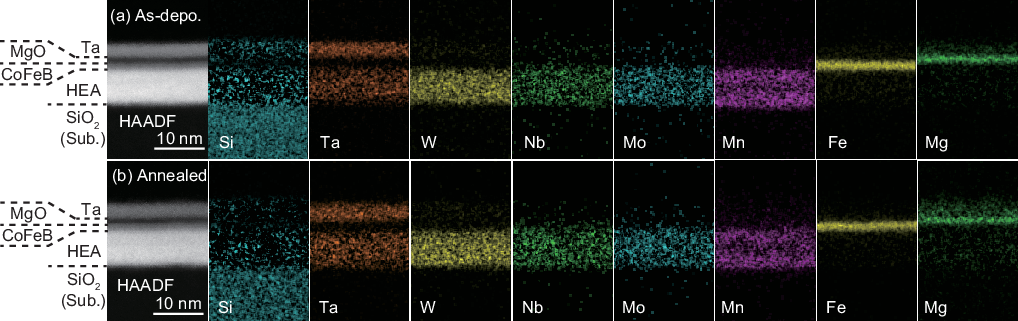}
\caption{Color mapping images for the energy dispersive X-ray spectra of (a) as-deposited and (b) annealed samples.}
\label{fig:EDX}
\end{center}
\end{figure*}
Energy-dispersive X-ray spectroscopy (EDX) mapping was also performed to confirm elemental distributions. The mapping images in Fig. \ref{fig:EDX} show the uniform distribution of elements within the designed layer region, and no significant elemental segregation was observed. These observations indicated that HEA films with a uniform single phase were fabricated.\\

\begin{figure}
\begin{center}
\includegraphics[clip,scale=0.7]{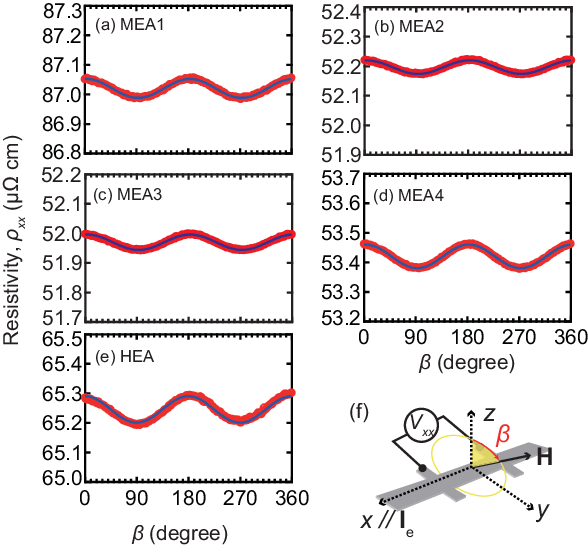}
\caption{(a--e) Angle-dependent resistivity change in HEA or MEA 5 nm $|$ CoFeB bilayer samples with different Mn--Nb--Mo--Ta--W compositions. All samples were annealed at 350 \degree C. Red thick and blue thin lines represent experimental data and fitting results, respectively. (f) Coordinate system for the spin Hall magnetoresistance measurements. The electron current (\textbf{I}$_\mathrm{e}$) flows along the $x$ direction. The direction of magnetic field (\textbf{H}) was changed in the $y$--$z$ plane with an angle $\beta$ by rotating the sample.}
\label{fig:SMRcurves}
\end{center}
\end{figure}
The film samples were patterned into 20-$\mu$m width stripes with voltage measurement terminals using photolithography and Ar ion dry etching.
Figs. \ref{fig:SMRcurves}(a--e) and \ref{fig:SMRcurves}(f) show the results of angle-dependent longitudinal resistivity ($\rho_{xx}$)  for the annealed samples with a thickness $t =$ 5 nm and a schematic illustration of the coordinate system for the measurement geometry, respectively. Here, $\rho_{xx} = R_{xx} \times w \times t / l = V_{xx} / I_\mathrm{e} \times w \times t / l$, where $w, l, I_\mathrm{e}, and V_{xx}$ are the stripe width, channel length for voltage measurements, charge current, and longitudinal voltage, respectively.
For $\rho_{xx}$ measurements, $I_\mathrm{e}$ was applied along the $x$ direction and the direction of magnetic field (\textbf{H}) was changed in the $y-z$ plane with an angle $\beta$ as depicted in Fig. \ref{fig:SMRcurves}(f), in which the contribution of anisotropic magnetoresistance was eliminated to ensure pure SMR measurements\cite{Nakayama2013}.
All the samples exhibit a resistance change with $\beta$ (Fig. \ref{fig:SMRcurves}(a --e)), and similar dependences were also observed for other thicknesses and as-deposited samples (not shown here).
Using the angular dependence of $\rho$, the SMR ratio is defined as follows:
\begin{equation}
\label{eq:ratio}
\mathrm{SMR \, ratio} = \frac{\Delta \rho_{xx}}{\rho_{xx}^{\beta=0}}\times 100 \, (\%),
\end{equation}
where $\Delta \rho_{xx}$ and $\rho_{xx}^{\beta=0}$ are resistivity change and the resistivity value, respectively, along the $x$ direction at $\beta = 0$ in the angular dependence of $\rho$.
\begin{figure}
\begin{center}
\includegraphics[clip,scale=0.7]{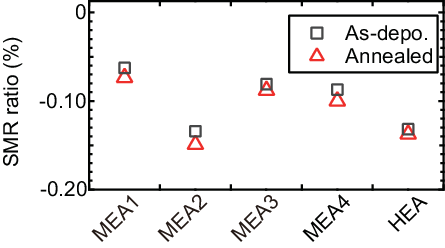}
\caption{SMR ratios for all the prepared samples. The layer thicknesses of MEAs and HEA were 5 nm for all data points. The black squares and red triangles represent the results for as-deposited and annealed samples, respectively.}
\label{fig:SMRcomp}
\end{center}
\end{figure}
The measured SMR ratios are summarized in Fig. \ref{fig:SMRcomp} for the as-deposited and annealed samples with $t =$ 5 nm. Note that the SMR ratios have a negative sign. MEA2 and HEA exhibited relatively large absolute values around -0.14\% compared with other samples, MEA2 exhibited relatively large SMR, possibly because it contained large amounts of heavy elements such as Ta and W. Contrarily, the HEA exhibits a notably high SMR comparable to MEA2, although it contained lesser amounts of heavy elements. This indicated high charge-to-spin conversion efficiency in HEAs.\\
The $t$ dependence of SMR ratio was also investigated for MEA4 and HEAs with relatively large $S^\mathrm{mix, ideal}$.
Based on a drift-diffusion model, the SMR ratio can be described as follows~\cite{Kim2016}:\\

\begin{equation}
\begin{split}
\label{eq:SMR}
\frac{\Delta \rho_{xx}}{\rho_{xx}^{\beta=0}} &\simeq -\theta^2_\mathrm{SH} \frac{\lambda_\mathrm{N}}{t} 
\frac{\tanh^2(\frac{t}{2\lambda_\mathrm{N}})}{1 + \xi}\\
&\left\{
\frac{g_\mathrm{R}}{1 + g_\mathrm{R} \coth(\frac{t_\mathrm{F}}{\lambda_{F}})}
- \frac{g_\mathrm{F}}{1 + g_\mathrm{F} \coth(\frac{t}{\lambda_\mathrm{N}})}
\right\},
\end{split}
\end{equation}
\begin{equation}
\label{eq:g_R}
g_\mathrm{R}\equiv 2 \rho_\mathrm{N}\lambda_\mathrm{N}\mathrm{Re}[G_\mathrm{MIX}],
\end{equation}
\begin{equation}
\label{eq:g_F}
g_\mathrm{F}\equiv\frac{(1 - P^2)\rho_\mathrm{N}\lambda_\mathrm{N}}
{\rho_\mathrm{F}\lambda_\mathrm{F}\coth(t_\mathrm{F}/\lambda_\mathrm{F})},
\end{equation}
\begin{equation}
\label{eq:xi}
\xi \equiv \frac{\rho_\mathrm{N}t_\mathrm{F}}{\rho_\mathrm{F}d},
\end{equation}
where $\theta_\mathrm{SH}$ is the spin Hall angle of the nonmagnetic layer. $\rho_\mathrm{N(F)}$, $\lambda_\mathrm{N(F)}$, and $t$($t_\mathrm{F}$) represent the resistivity,  spin diffusion length, and layer thickness of the nonmagnetic (ferromagnetic) layer, respectively. $G_\mathrm{MIX}$ is the so-called spin mixing conductance~\cite{Brataas2000,Weiler2013,Kim2014} that defines the absorption of the transverse spin current impinging on the NM$/$ferromagnetic layer interface~\cite{Stiles2002}.
$\theta_\mathrm{SH}$ and $\lambda_\mathrm{N(F)}$ were determined using Eqs. (\ref{eq:SMR})--(\ref{eq:xi}), which provide qualitative comparisons of the spin conversion efficiencies for the studied materials. 
\begin{figure}
\begin{center}
\includegraphics[clip,scale=0.7]{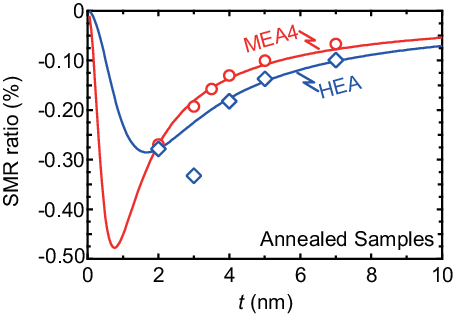}
\caption{Thickness dependence of SMR ratios. Red circles and blue diamonds represent datapoints for MEA4 and HEA samples, respectively. Solid lines represent fitting results based on Eqs. (\ref{eq:SMR})--(\ref{eq:xi}). Note that the HEA datapoint at $t =$ 3 nm was neglected for fitting because no reasonable fitting was achieved with it.}
\label{fig:SMRthick}
\end{center}
\end{figure}
Fig. \ref{fig:SMRthick} shows the $t$ dependence of SMR ratios for the annealed MEA4 and HEA. With decreasing $t$, the absolute value of SMR ratio increased. For fitting, $\rho_\mathrm{N}$ was kept constant at 40.24 and 56.72 $\mu \Omega$ cm for MEA4 and HEA layers, respectively; these values were obtained from the experimental values of the $t$ dependence of $R_{xx}$ in the layered samples. 
In other words, linear fittings were performed for the $1/R_{xx}$ of MEA4 and HEA samples as a function of $t$, and the slope values were used for determining the $\rho_\mathrm{N}$ of MEA4 and HEA layers\cite{Masuda2020}.
The values of $\rho_\mathrm{F}$, $G_\mathrm{MIX}$, $\lambda_\mathrm{F}$, $P$, and $t_\mathrm{F}$ were also kept constant at 160 $\mu \Omega$ cm, $1 \times 10^{15}~\Omega^{-1} \mathrm{m}^{-2}$, 1 nm, 0.72, and 1.5 nm, respectively; these were obtained from the literature\cite{Liu2015,Masuda2020,Saito2020}. Note that the data point at $t =$ 3 nm of the HEA was neglected for fitting because no reasonable fitting result was achieved.
\begin{table}
\begin{center}
\caption{
Fitting results shown in Fig. \ref{fig:SMRthick} and reported values of $\theta_\mathrm{SH}$ and $\lambda_\mathrm{N}$ in the literature. A review article can also be found for other material systems\cite{Hoffmann2013}.
}
\label{t:fitting}
\begin{ruledtabular}
\begin{tabular}{c ccc}
Material system & $|\theta_\mathrm{SH}|$ & $\lambda_\mathrm{N} (nm)$ & Reference\\
\hline
MEA4$/$CoFeB & 0.14 $\pm$ 0.037 & 0.33 $\pm$ 0.21 & this work\\
HEA$/$CoFeB & 0.12 $\pm$ 0.002 & 0.72 $\pm$ 0.03 & this work\\
\hline
$\beta$-W$/$CoFeB & 0.33$\pm$0.06 & -- & \cite{Pai2012}\\
$(\alpha+\beta)$-W$/$CoFeB & 0.18$\pm$0.02 & -- & \cite{Pai2012}\\
$\alpha$-W$/$CoFeB & $<$0.07 & -- & \cite{Pai2012}\\
$\beta$-W$/$CoFeB & 0.27 & 1.26 & \cite{Kim2016}\\
Cu$_{0.995}$Bi$_{0.005}$ & 0.021$\pm$0.006 & -- & \cite{Niimi2012}\\
Cu$_{0.95}$Ir$_{0.05}$$/$Co & 0.043 & 1.3 $\pm$ 0.1& \cite{Masuda2020}\\
Pt$/$Y$_3$Fe$_5$O$_{12}$ & 0.04 & 2.4 & \cite{Nakayama2013}\\
Pt$/$Y$_3$Fe$_5$O$_{12}$ & 0.11$\pm$0.08 & 1.5$\pm$0.5 & \cite{Althammer2013}\\
Pt$/$CoFeB & 0.14 & 1.1 & \cite{Magni2022}\\
\end{tabular}
\end{ruledtabular}
\end{center}
\end{table}

The fitting results for $\theta _\textrm{SH}$ and $\lambda_\mathrm{N}$ are summarized in Table \ref{t:fitting}, which also shows previously reported values for materials such as typical spin Hall materials. The values of $\theta_\textrm{SH}$ are 0.14$\pm$ 0.037 and 0.12 $\pm$ 0.002 for the MEA4 and the HEA, respectively, which are comparable to that of a Pt$/$CoFeB system\cite{Magni2022}. The average atomic numbers for MEA4 and HEA are 55.1 and 50.5, respectively, which are much smaller than the atomic number of 78 for Pt.
Compared with some emerging spin Hall materials, such as $\beta$-W\cite{Pai2012,Kim2016} and topological Bi-compounds\cite{Mellnik2014,DC2018,Khang2018}, $\theta_\mathrm{SH}$ values for HEA and MEA4 are moderately large. However, the stable bcc phase of the HEA and MEAs has a practical merit for spintronic applications, such as memory devices and sensors, in which magnetic tunnel junctions (MTJs) with MgO tunneling barrier and CoFeB ferromagnets are essential to obtain readout signals. Moreover, bcc materials are suitable for the stacking structures of MgO--CoFeB-based layered samples to obtain sufficient signals from the MTJs\cite{Djayaprawira2005,Yuasa2005}.
Thus, the as-fabricated HEA and MEAs are attractive as spin Hall material systems.

\section{Conclusions}
Herein, the crystal structures and SMR of layered structures composed of an HEA/MEA film and a CoFeB film were investigated.
XRD and TEM results indicated that the MEA and HEA layers exhibited a single bcc phase. EDX mapping also showed uniform distributions of elements in the HEA.
The SMR was observed for all HEA and MEA compositions.
The layer thickness dependence of SMR was analyzed based on the drift diffusion model for the HEA and MEA4, which revealed $\theta_\mathrm{SH}$ of 0.12 $\pm$ 0.002 and 0.14 $\pm$ 0.037, respectively.
These values of $\theta_\mathrm{SH}$ were comparable with that of Pt, which is a typical heavy element with a large $\theta_\mathrm{SH}$.
The HEA and MEAs are thus attractive from the viewpoint of spin Hall materials with a stable bcc phase.

\section*{Acknowledgement}
This work was partly supported by KAKENHI (JP21K18180) and the GIMRT program of the Institute for Materials Research (IMR), Tohoku University (Proposal No. 202012 CRKEQ 0417). TK and KT would like to thank Yue Wang of Lanzhou University, China, for fruitful discussion during a term for her COLABS program in Tohoku University. TK would like to thank Issei Narita of the IMR for technical support.

\bibliographystyle{aipnum4-2}
\bibliography{
d:/Documents/BibTex/SpinOrbitronics.bib,
d:/Documents/BibTex/HighEntropyAlloys.bib,
d:/Documents/BibTex/Theory.bib
}
\end{document}